\documentclass[iop]{emulateapj}

\newcommand{\uvceti}{UV~Ceti}

\usepackage{amsmath}
\usepackage[caption=false]{subfig}
\usepackage{natbib}
\usepackage{graphicx}
\usepackage{longtable}

\begin{document}
\title{154 MHz detection of faint, polarized flares from UV Ceti}

\author{
C. R. Lynch,$^{1,2}$
E. Lenc,$^{1,2}$
D. L. Kaplan,$^{3}$ 
Tara Murphy,$^{1,2}$
G. E. Anderson$^{4}$}
\affil{$^1$ Sydney Institute for Astronomy, School of Physics, The University of Sydney, NSW 2006, Australia}
\affil{$^2$ ARC Centre of Excellence for All-sky Astrophysics (CAASTRO)}
\affil{$^3$ Department of Physics, University of Wisconsin--Milwaukee, Milwaukee, WI 53201, USA}
\affil{$^4$ International Centre for Radio Astronomy Research, Curtin University, GPO Box U1987, Perth, WA 6845, Australia}

\begin{abstract}
We have detected four flares from \uvceti\ at 154 MHz using the Murchison Widefield Array. The flares have flux densities between 10--65\,mJy --- a factor of 100 fainter than most flares in the literature at these frequencies --- and are only detected in polarization. The circular polarized fractions are limited to $>27$\% at 3$\sigma$ confidence and two of the flares exhibit polarity reversal. We suggest that these flares occur periodically on a time scale consistent with the rotational period of UV Ceti. During the brightest observed flare, we also detect significant linear polarization with  polarization fraction $>18$\%. Averaging the data in 6-minute, 10\,MHz frequency bins we find that the flux density of these flares does not vary over the 30\, MHz bandwidth of the Murchison Widefield Array, however we cannot rule out finer time-frequency structure. Using the measured flux densities for the flares, we estimate brightness temperatures between $(10^{13}-10^{14})\,$K, indicative of a coherent emission mechanism. The brightness temperature and polarization characteristics point to the electron cyclotron maser mechanism. We also calculate the flare rates given our four observed flares and compare them to flare rates for the set of M dwarf stars with known 100--200\,MHz flares. Our measurement is the first for flares with intensities $<100$\,mJy at 100-200 MHz.


\end{abstract}

\keywords{polarization -- radiation mechanisms: Nonthermal -- radio continuum: stars -- stars: low-mass -- stars: flare}

\section{Introduction} \label{sec:intro}

Flaring is a common characteristic of magnetically active stars. These flares are driven by reconnection events that convert magnetic energy to  particle kinetic energy and bulk plasma motion. During these events all stellar atmospheric layers are affected and brightening is observed across the full electromagnetic spectrum \citep{Benz:2010}. M dwarf stars have flare energies up to 10$^3$ times that of the Sun \citep{Haisch:1989}. These objects are near-to-fully convective and have kilo-Gauss magnetic fields covering large fractions of the stellar disc \citep{Johns-Krull:1996}. 

\subsection{Radio flares from M dwarfs}
Radio flares from  M dwarf stars are dominated by coherent emission at frequencies less than 5\,GHz. This radiation originates from non-thermal electron populations excited by high-frequency waves. Coherent radio emission provides information about the acceleration process involved in generating flares and the characteristics of the magnetospheric plasma, such as density and magnetic field strength \citep{Benz:2010}.
 
Early radio observations targeted optically-active M dwarf stars and were carried out at meter wavelengths using single-dish radio telescopes \citep{Lovell:1964a, Lovell:1969, Higgins:1968, Slee:1963,  Spangler:1974a, Spangler:1976, Nelson:1979}. These observations were sensitive to flares with intensities greater than several hundred millijanskys. In general these flares had durations ranging from 0.5 to 3 hours, large negative spectral indices, high circular and in some cases linear \citep{Spangler:1974a} polarization,  and extremely high brightness temperatures. Simultaneous optical and radio observations found an association between flares in these two wavebands, where optical flares were followed by radio flares with delays ranging from a few minutes \citep{Lovell:1974} to 30 minutes \citep{Spangler:1976}. This correlation led previous authors to assume the radio flares are analogous to Solar radio emission at meter wavelengths (e.g., burst types II or IV; \citealt{Bastian:1990a}).

Distinguishing between terrestrial radio frequency interference and stellar flare emission is difficult using single dish telescopes (even with auxiliary feed systems such as that used by \citealt{Spangler:1974b}) and relies on observations of these flares over multiple wavelengths. In contrast, interferometric observations can easily distinguish between these types of radio emission. Yet there have been only a few such observations of M dwarf stars at meter wavelengths, with only two detections of low-intensity ($\leq$100 mJy) flares \citep{Kundu:1988a, Davis:1978, Gudel:1989, Jackson:1990, Boiko:2012, Crosley:2016}. Bright flares with intensities ranging from hundreds of millijanskys to Jansky-level have yet to be detected with interferometric observations at MHz frequencies.  

\begin{figure}[t!]
\centering
 \includegraphics[width=\columnwidth]{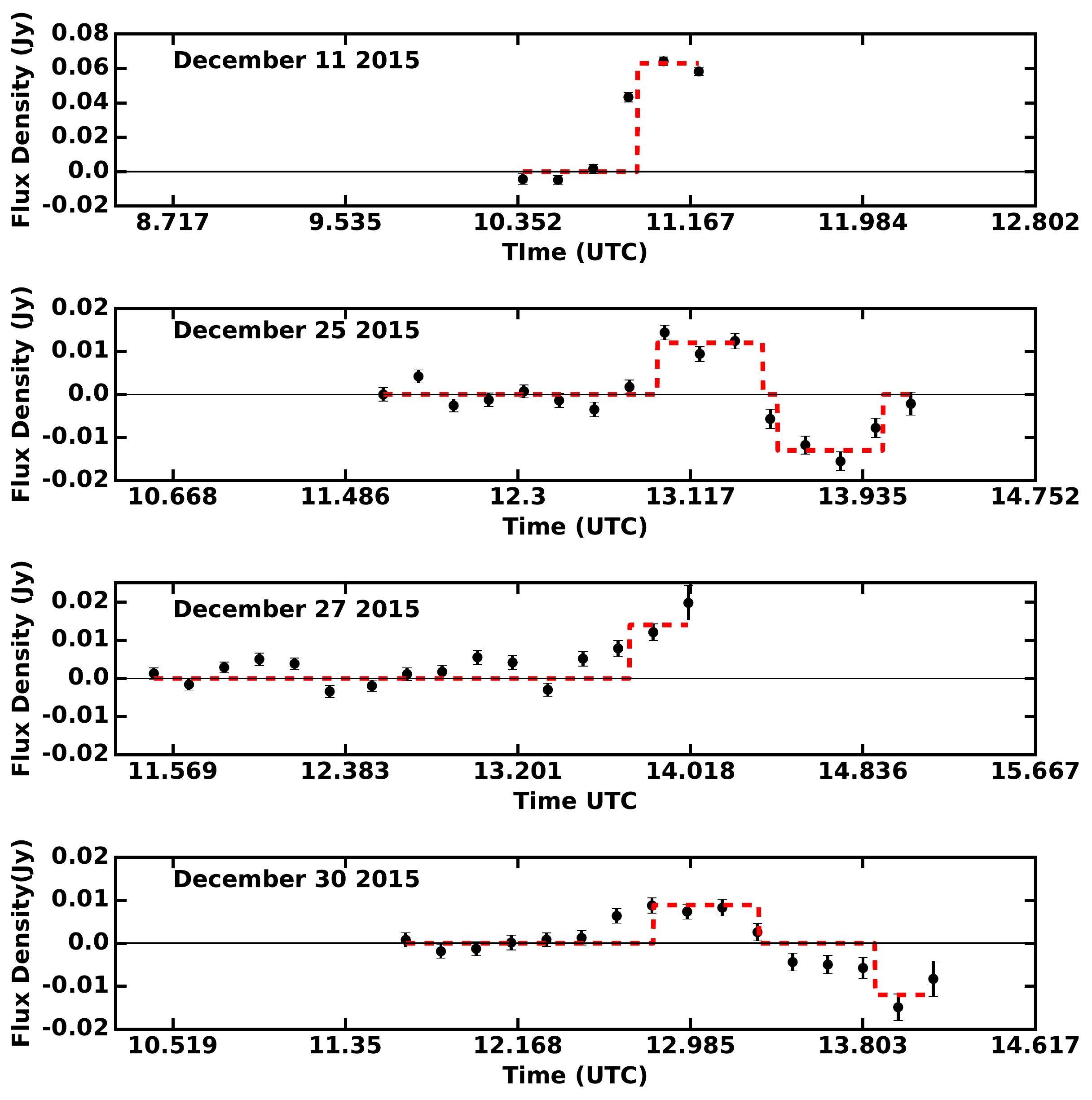}
 \caption{10-minute Stokes V light curves for the four observational epochs.  A positive flux density corresponds to right-handed circular polarization and negative corresponds to left-handed circular polarization. Overlaid is the best-fit flare model for each epoch (red-dashed line).}
 \label{fig:LC}
\end{figure}

Recently, we targeted \uvceti\ using the Murchison Widefield Array \citep[MWA;][]{Tingay:2013} to confirm previous bright flare detections  at 100--200\, MHz. We detected four circularly polarized, low-level flares from  \uvceti. This is the first detection of meter-wave flare star emission using modern low-frequency radio interferometers.

\section{Observations and Data Reduction}

We observed \uvceti\ with the MWA over four observing sessions during 2015~December, for a total of 8.8\,hours. All observations used a 30.72\,MHz bandwidth centered at 154\,MHz, with 40\,kHz channels and 0.5\,s integrations. During each observing session data were separated into 2\,min individual observations, with a slightly different pointing center used for each in order to track \uvceti\ while retaining the sensitivity of the MWA (offsets between the pointing center and \uvceti\ were typically a few degrees, compared to the  primary beam FWHM of $\sim 30\degr$). For the first observation two MWA receivers were off-line removing 8 tiles each, which led to a slightly elongated synthesized beam shape.  



Bandpass and gain calibration was performed using the Real-time System (\textsc{RTS}; \citealt{Mitchell:2008,Ord:2010}) software package. We used a sky model centered on the inner $20\degr$ region of the primary beam pointing and taking an ensemble of the 50 brightest sources taken from the MWA Commissioning Survey \citep{Hurley-Walker:2014}. Visibility data from baselines shorter than 50\,$\lambda$ were flagged to avoid contamination from diffuse structure. Full spectral resolution cubes were generated in all four Stokes parameters utilizing both natural weighting and a robust weighting of zero. Full Stokes continuum maps were also created from the spectral cubes by averaging over the frequency channels. We measure the RMS in the center of each image following the method of \citet{Swinbank:2015}. The 1$\sigma$ RMS sensitivity of a single Stokes I, 2-minute image is 90\,mJy and in Stokes V is 9\,mJy. The 1$\sigma$ RMS sensitivity of a 10-minute Stokes I image is  90 mJy and for Stokes V is 5.3 mJy. Because the circularly polarized images are not sidelobe or classical confusion limited (cf.\ \citealt{Franzen:2016}) we are able to achieve much better sensitivity in these images. 

Upon inspection of the resulting full Stokes continuum images, polarization leakage at the 7\%, 2\% and 0.1\% levels was noted for Stokes Q, U, and V, respectively. The leakage results from inaccuracies in the primary beam model \citep{Sutinjo:2015} which increase at low elevations but are relatively constant over the $2\degr$ region centered around \uvceti. To mitigate the effect of polarization leakage we perform a simple subtraction of the leaked proportion of Stokes I from the remaining Stokes images. A single leakage fraction is determined for each snapshot and each Stokes parameter by measuring the fractional polarization against a bright field source assumed to be unpolarized. For each frequency channel of a given snapshot, the leaked proportion of the Stokes I image was subtracted from each of the Stokes Q, U, and V images. With this method, the resulting leakage in the images was reduced to below $0.1\%$ for all of the Stokes parameters.

\section{Characteristics of the Radio Emission}
To search for flare emission from \uvceti\ we measured the peak flux at the pixel location of the source in each of the Stokes I and Stokes V continuum maps and created both 2-minute and 10-minute averaged light curves for each epoch.  Visual inspection of the light curves quickly revealed flaring activity (see Figure \ref{fig:LC}). Table \ref{tab:summary} lists the measured Stokes V flare intensities during each observation as well as the measured central timing of each flare. 

\begin{figure*}[ht!]
\centering
 \includegraphics[scale=0.6]{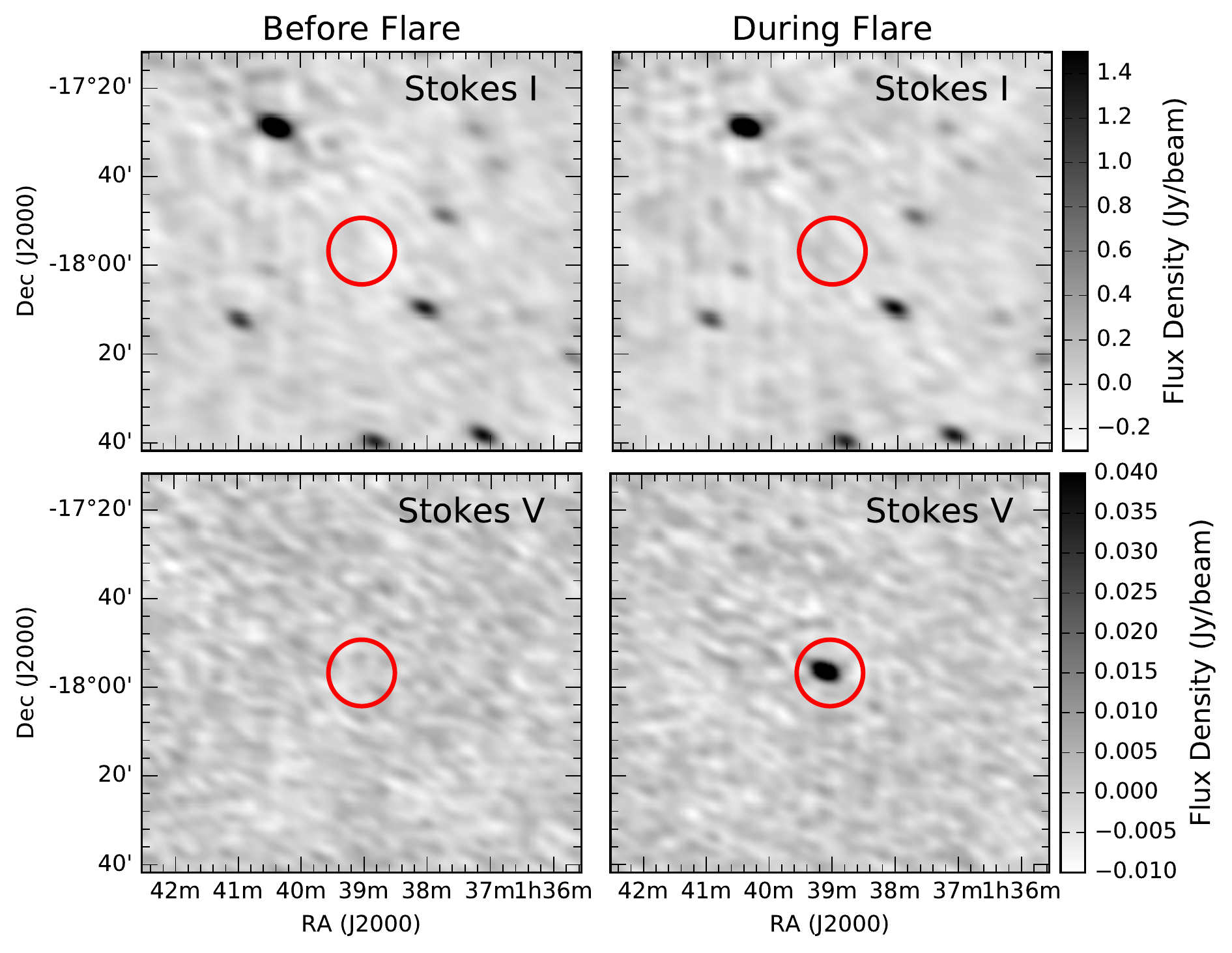}
 \caption{Stokes I (top) and Stokes V (bottom) images during 2015 December 11 observation, averaged over the 30 minutes before (left) and during (right) the flare. The 1$\sigma$ RMS in the Stokes I image is 85 mJy and in the Stokes V image 2 mJy.}
 \label{fig:array}
\end{figure*}

\subsection{Polarization}

Due to their greater sensitivity, we only detect the flares in the polarized images. This is illustrated in Figure \ref{fig:array}, where we show both the Stokes I (top) and Stokes V (bottom) continuum maps from the 2015 December 11 observation, averaged over the 30 minutes minutes before (left) and during (right) the flare. There is a clear Stokes V detection at the location of \uvceti\ during time of the flare with no corresponding source in the Stokes I map. Taking the 3$\sigma$ limit on the Stokes I emission during the flare, we place a lower limit on the fractional circular polarization  of $\geq 0.27$. This is the first case in which stellar flare emission has been detected in polarized emission without a total intensity counterpart. It highlights the importance of using the polarized images to look for low-intensity transient events with the MWA. 

In each epoch we detect a single right-handed circularly polarized flare. However for 2015 December 25 and 30, we also detect a left-handed flare immediately following the right-handed flare. It is possible that all of the observed right-handed flares are followed by left-handed flares. We are unable to confirm this, as our observations of \uvceti\ end mid-flare during both the 2015 December 11 and 27 epochs.  Similar polarity reversal is observed in some brown dwarfs between  4 - 8\,GHz \citep{Hallinan:2007, Lynch:2015}. In these previous cases, the polarization behaviour is attributed to a highly inclined ($\sim$ 90 degrees) magnetic axis. Such an orientation of the magnetic axis allows for observations of flaring activity from both magnetic poles \citep{Route:2016}. The observation of both polarities occurring in \uvceti\ similarly suggests a high inclination angle for the magnetic axis.

We searched for linearly polarized emission during the brightest  of the observed flares. To improve the prospects of detection, the robust-weighted image cubes were integrated in time separately over the off and on period. Rotation Measure (RM) Synthesis \citep{Brentjens:2005} was performed over the integrated Stokes Q and U cubes. In the resulting RM cube for the off period, no detection was made. However, in the RM cube for the on period, a single polarized source was detected with a flux density of 42\,mJy\,beam$^{-1}$\,RMSF$^{-1}$ (polarization fraction $>0.18$) coincident with \uvceti, indicating that these flares are elliptically polarized.

\begin{figure}[!t]
\centering
 \includegraphics[width=\columnwidth]{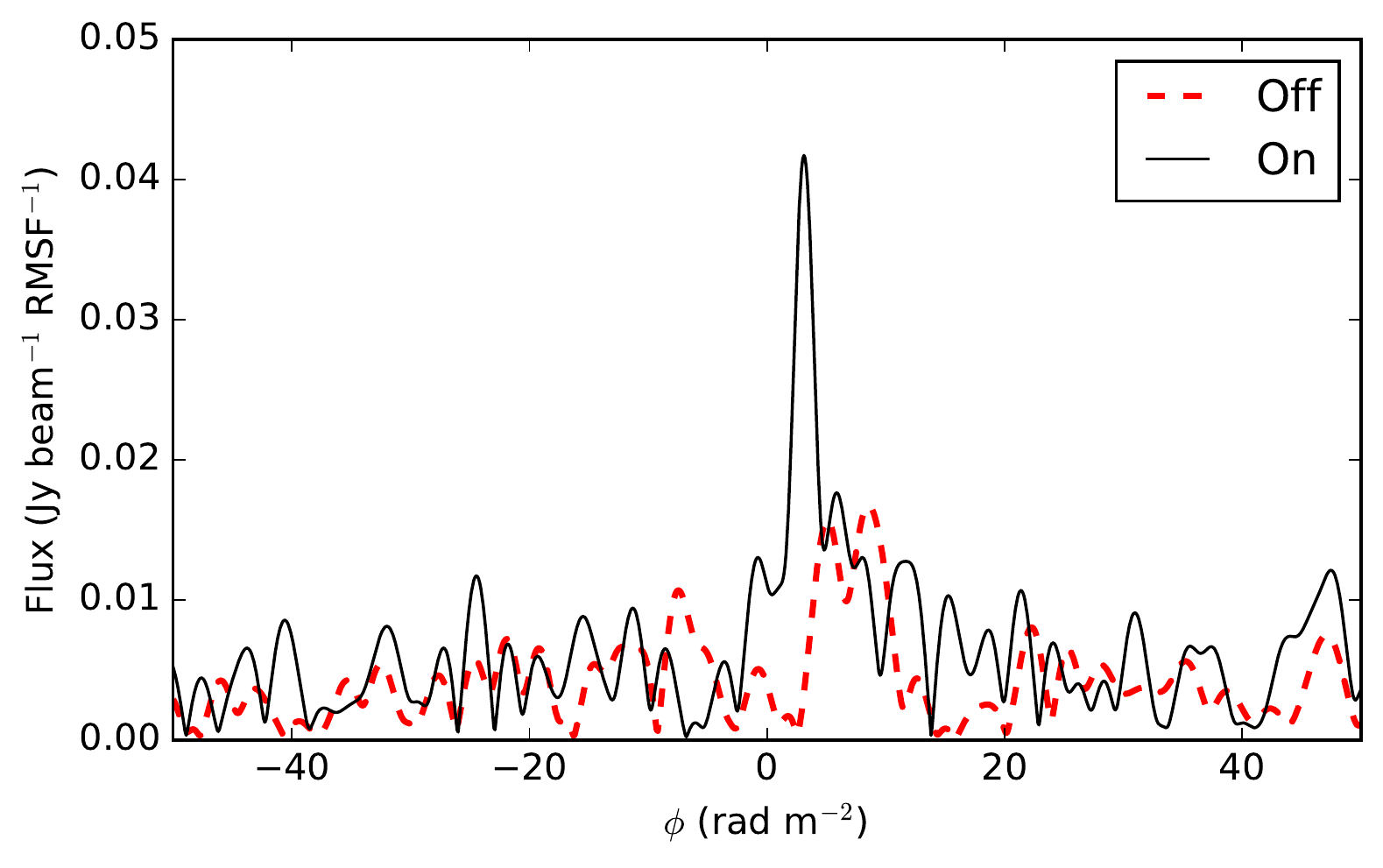}
 \caption{Faraday dispersion function of \uvceti\ at $154$\,MHz during the off and on period of the 2015 December 11 flare. During the on period the polarized intensity peaks at $42$\,mJy\,beam$^{-1}$\,RMSF$^{-1}$ at $\phi=+3.0$\,rad\,m$^{-2}$. The noise is $4.1$\,mJy\,beam$^{-1}$\,RMSF$^{-1}$}
 \label{fig:fdf}
\end{figure}

To remove the effect of sidelobes in Faraday space, the RM cube for the detection was deconvolved using RM CLEAN \citep{Heald:2009}.  To estimate and correct for the ionospheric component of Faraday rotation during the observation, the software tool \textsc{RMextract}\footnote{\url{https://github.com/maaijke/RMextract}} was used to estimate an ionospheric component to the RM of $-1.9$\,rad\,m$^{-2}$. The Faraday dispersion function for \uvceti\, is shown in Figure \ref{fig:fdf}. We estimate the galactic foreground contributes RM$\sim$0.001 rad m$^{-2}$ to our measurement using data from the angularly nearby pulsar PSR B0149-16 \citep{Mitra:2016}. Neither the intervening ionosphere or interstellar medium can account for the measured RM and so it must be inherent to the atmosphere of the star.

For stars, linear polarization is usually assumed to be unobservable due to Faraday rotation in the overlying plasma \citep{Dulk:1985}. Using the observed linear polarization we can place a constraint on the Faraday rotation in the atmosphere of \uvceti. The measured RM of 3.0 rad m$^{-2}$ is associated with a total Faraday rotation at 154\,MHz of 12 radians.

\subsection{Frequency structure}

To analyze the spectral properties of the flare emission at 154\,MHz, we made 5 spectral cubes of the flare on 2015 December 11, using 6 minute, 10 MHz bin averaging. Each image has a measured RMS of 9 mJy. Over the full 30-MHz bandwidth, the emission does not vary with time or frequency within the uncertainty of the measurements. We cannot rule out finer time-frequency structure as we do not have the sensitivity to look in smaller time-frequency  bins.

\subsection{Periodicity}\label{sec:period}

Both magnetic chemically peculiar stars \citep[e.g.][]{Trigilio:2000} and brown dwarfs \citep[e.g.][]{Wolszczan:2014} are observed to exhibit highly circularly polarized bursts that are periodic on time scales of their rotation. Similar periodicity might be expected for the 154\, MHz flares we observe from \uvceti\ . In order to determine if the flares are periodic, we first measure the mid-point time and amplitude for each flare by fitting a top-hat function and assume a constant width of 30 minutes. To constrain the uncertainties in the measured flare times and amplitudes we simulated light curves using the best fit model and added a noise component equal to the RMS. The uncertainty was taken to be the standard deviation of the distribution of the fit amplitudes and times for the simulated data. The measured amplitudes and flare times are given in Table \ref{tab:summary} and best-fit top-hat models are overlaid for each epoch in Figure \ref{fig:LC}.

The rotational period of \uvceti\ has yet to be determined, but the projected rotational velocity, $v \sin(i)\sim$30 km s$^{-1}$, places a limit on the rotational period of $\leq$6.5 hours \citep{Kervella:2016}.  We  check if we can fit the flare times for a period consistent with this limit using the right-handed flare times and minimizing the phase difference between the measured flare times and model flare times given a set of trial periods, $P\ \leq$6.5 hours. We find that data is fit by several periods consistent with this limit and indicates that the observed variability may be due to the rotation of the star. Follow up observations of these flares will allow for better constraints to be placed on the flare period.


\begin{figure*}[ht!]
\centering
 \includegraphics[scale=0.6]{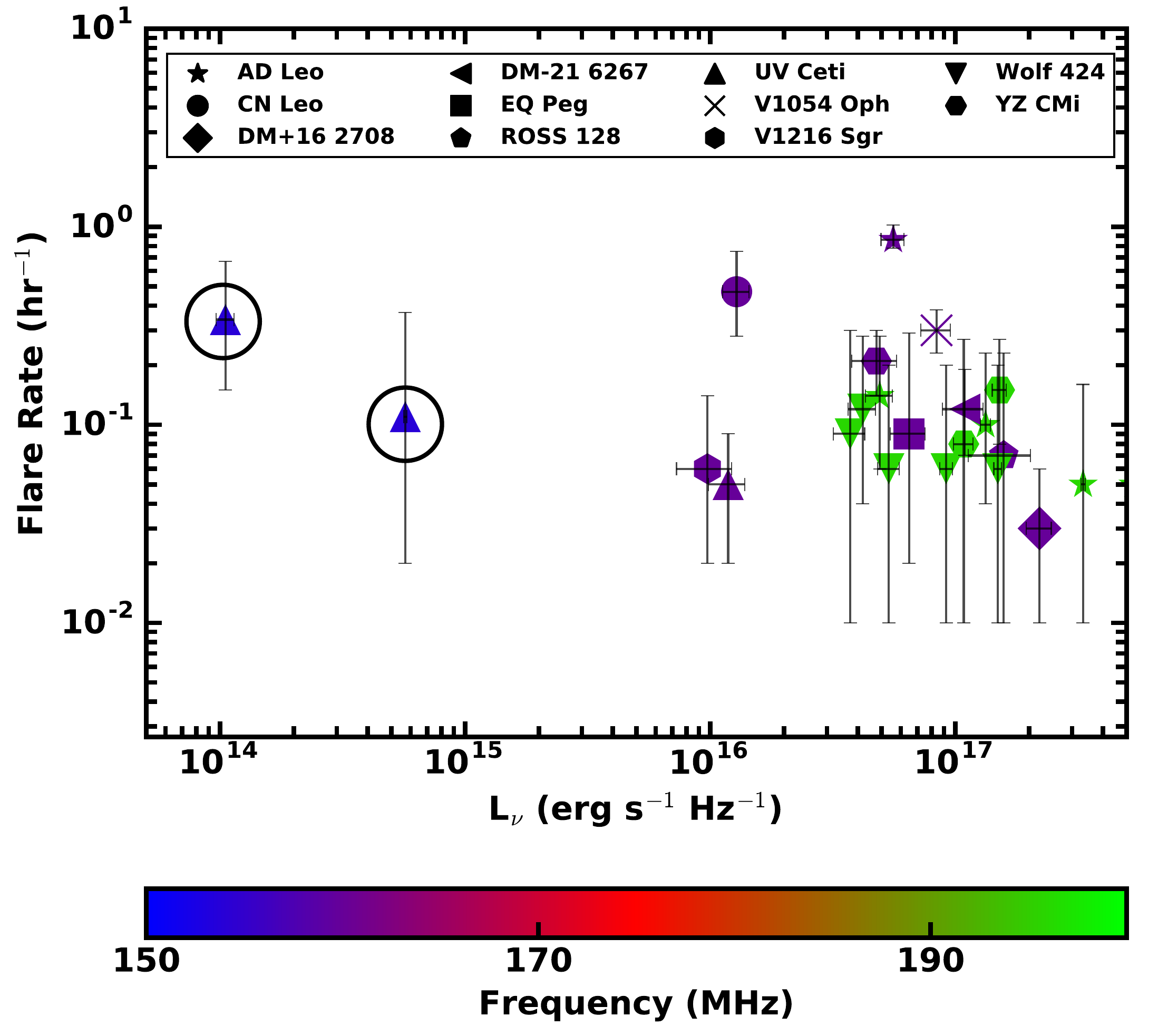}
 \caption{Flare rates as a function of spectral luminosity for M dwarf stars with observed flares at 100 - 200\,MHz \citep{Spangler:1974b, Spangler:1976, Nelson:1979}.The color is scaled to the observational frequency. The flare rates associated with the flares detected in our MWA observations are highlighted by a black circle.}
 \label{fig:rates}
\end{figure*}

\subsection{Brightness Temperature}
To determine whether the flare emission is driven by a coherent or incoherent radiation mechanism, we calculate the brightness temperature of the emission at 154\,MHz. Using the expression in \citet{Dulk:1985} we can express the flux density, $S_{\nu}$ as:
\begin{equation}
S_{\nu} = 2k_BT_b\left(\frac{\nu}{c}\right)^2\left(\frac{l}{d}\right)^2
\end{equation} where  $k_B$ is  Boltzmann's constant, $T_b$ is the brightness temperature of the emission at a frequency $\nu$, $c$ is the speed of light, $l$ is the size of the source region,  and $d=2.68$\,pc \citep{vanAltena:1995} is the distance to \uvceti.  

We suggest that the variability of the emission may be the result of a rotating region of persistent emission. If the emission is highly beamed, we will only detect the emission when the source region crosses our line of sight. Therefore, the combination of beaming and rotation can be used to measure the size of this region. The size of the region will be related to the rotational velocity and duration $\Delta t$ of the emission, $l$ = $\Delta t\,v$. Using the projected rotational velocity of 30 km s$^{-1}$ and the duration of 30 minutes, we constrain the size to be $5.4\times 10^7\,$m. We calculate brightness temperatures for the observed flares to be between $(0.3 - 2.1)\times10^{14}$\,K. 

The brightness temperature of  incoherent synchrotron emission is limited to $T_b\ \leq$ 10$^{12}$ K by inverse Compton scattering \citep{Dulk:1985}. Brightness temperatures greater than this are explained using coherent emission mechanisms, such as plasma emission or electron cyclotron maser (ECM) emission. Even if we assume that the emission is coming from the full corona of \uvceti\, whose size is measured to be $5.6\times10^8$\,m \citep{Benz:1998},  we calculate a range of brightness temperatures, $T_b$ = $(0.9 - 6.2)\times 10^{13}$\,K, still consistent with coherent emission.

\section{Multi-wavelength coverage}

To investigate whether a correlation can be made between X-ray flares and the observed 154\,MHz flares, we searched the archives of several X-ray telescopes. \uvceti\ is monitored by the Swift Burst Alert Telescope \citep[BAT;][]{Krimm:2013} and the Monitor of All-sky X-ray Image \citep[MAXI;][]{Matsuoka:2009} high energy missions. Inspection of the daily averaged light curves of \uvceti\ during the MWA observations show no significant variability that can be attributed to an associated high energy flare. Optical and low-frequency radio correlations have been observed \citep{Spangler:1974b}, however we were unable to find any optical data covering \uvceti\ during our observations.

\section{Discussion}
\subsection{Emission Mechanism}

The high brightness temperatures indicate that the observed 154\, MHz flares are produced by a coherent emission mechanism. Two emission mechanisms are suggested to produce coherent stellar emission: ECM and plasma emission. ECM is considered to be more efficient than plasma emission \citep{Zautsev:2005},  yet due to the possible re-absorption of ECM in the stellar atmosphere, several authors argue for plasma emission \citep[e.g.][]{Stepanov:2001, Osten:2006}. 

To differentiate between these two emission mechanisms the observed polarization is key. Elliptical polarization of astronomical sources is relatively rare and is most likely the result of radiation generated in x-mode at an angle nearly perpendicular to the magnetic field, in a low density region such that the ratio of the electron plasma frequency, $\nu_p\sim9\ \text{kHz}\ \sqrt[]{n_e}$, to the electron cyclotron frequency, $\nu_c\sim2.8\ \text{MHz}\ (B)$, is small ($\nu_p$/$\nu_c\, < 1$) \citep{Dulk:1994}. From theoretical studies, ECM is completely polarized in the sense of x-mode, while plasma emission will be polarized in the sense of o-mode \citep{Melrose:1993}. Thus the observation of elliptically polarized coherent flares from \uvceti\ are indicative of ECM emission.

Due to the expected re-absorption of fundamental ECM emission, it is likely that only second harmonic x-mode emission will be observed from \uvceti. Second harmonic ECM emission at 154\, MHz corresponds to a local magnetic field strength of 28\, Gauss. The requirement that $\nu_p$/$\nu_c\, << 1$ puts a constraint on the local electron density of $n_e\lesssim$ 7$\times\ 10^7$\, cm$^{-3}$.  These values are consistent with the magnetospheric model for \uvceti\ from \citet{Benz:1998}, who estimate a field strength B $=$ 15 - 130\, Gauss and electron density $n_e <\, 10^8$\, cm$^{-3}$.

\subsection{Flare Rates}

To characterise the flare rates at 100--200\,MHz , we collect records of M dwarf flares that occured within this frequency range \citep{Spangler:1974b, Spangler:1976, Nelson:1979}. To be consistent with publications reporting only the mean flux density, we grouped individually reported flares and take the mean of the reported flux densities. We also calculate flare rates for our observed flares by grouping the three low-level flares ($<15$\, mJy) and treating the bright flare as an isolated event. The uncertainties for all the flare rates were calculated using the Poisson confidence interval approximation from \citet{Gehrel:1986}. These uncertainties are large due to the  short duration of the observations, with most searches lasting $\leq 20$\,hours in total. 

Figure \ref{fig:rates} shows the  calculated flare rates as a function spectral luminosity; the symbols represent different stars and color represents the observing frequency.  We highlight our MWA observations using a black circle. These are the first flare rate measurements for low-intensity ($\leq 100$\,mJy) flares at 100--200\,MHz.  For many objects there is a single flare rate reported in this frequency range, making it impossible for us to characterize a trend with luminosity. For objects with more than one reported flare rate, we find that the rates are consistent with being constant with luminosity given the large uncertainties in the flare rates.

\section{Summary}
Flaring activity is a common characteristic of magnetically active stellar systems. Observations of radio flare emission place constraints on the physical parameters of the stellar magnetospheric plasma. While radio flares at 1--4\,GHz are well studied for M dwarf stars, flares occurring at frequencies $<1 $\,GHz are not well observed for this group of magnetically active objects.  

We detect four flares from the M dwarf \uvceti\ at 154\,MHz using the MWA. These dim flares are only detected in polarized images, which have an order of magnitude better sensitivity than the total intensity images. This highlights the utility of using  polarization images to detect low level emission in confusion limited images. The circular polarization is observed to change polarity for two of the observed flares. Additionally we detect linear polarization during the brightest flare, indicating that the flares are elliptically polarized. We find that the flares are potentially periodic on the same time-scale as the rotation of \uvceti. Future observations are needed in order to better constrain the period as several different periods are possible. The calculated brightness temperature indicates a coherent emission mechanism and the elliptical polarization uniquely identifies the emission to be ECM. The flares have constant flux density across the full 30\,MHz bandwidth, however finer variability cannot be ruled out due to our low signal-to-noise. Lastly, using these detected flares, we measure the first flare rates for flares with intensities $< 100$\,mJy at 154\,MHz. Due to the small number of 100--200\,MHz flares measured for M dwarfs in general, the trend of the flare rate with luminosity is not well constrained. 

\begin{deluxetable*}{l c c c c c c c c c c}[!ht]
\tabletypesize{\small}
\tablewidth{0pt}
\tablecaption{\label{tab:summary}Summary of MWA Observations of \uvceti}
\tablehead{
\colhead{Date} & \multicolumn{2}{c}{Time (UTC)}  & \multicolumn{2}{c}{Elevation (deg)} 
 & \colhead{Number} & \colhead{Flare Amplitude} & \colhead{Polarity} & \colhead{Flare Time}\\
 & \colhead{Start} & \colhead{End} & \colhead{Start} & \colhead{End} &\colhead{of Antennas}& \colhead{(mJy)} & &\colhead{(UTC)}
}
\startdata 
2015-12-11 & $10^{\rm h}23^{\rm m}$ & $11^{\rm h}21^{\rm m}$ & 58.6 & 71.0 &  105 & $63\pm1$   &  RH & $11^{\rm h}10^{\rm m}42\pm40^{\rm s}$\\
2015-12-25 & $11^{\rm h}32^{\rm m}$ & $14^{\rm h}10^{\rm m}$ & 81.1 & 53.8 &  119 & $12\pm1$   & RH  &$13^{\rm h}07^{\rm m}13\pm50^{\rm s}$\\
		   &					  &				           &	     &         &             &  $-13\pm1$ & LH  &13$^{\rm h}$41$^{\rm m}$33$\pm$40$^{\rm s}$\\
2015-12-27 & $11^{\rm h}31^{\rm m}$ & $14^{\rm h}03^{\rm m}$ & 81.2 & 53.5 &  119 & $14\pm2$   & RH  &$14^{\rm h}01^{\rm m}55\pm50^{\rm s}$\\
2015-12-30 & $11^{\rm h}33^{\rm m}$ & $14^{\rm h}05^{\rm m}$ & 80.5 & 50.5 &  119 & $9\pm1$     & RH  &$12^{\rm h}59^{\rm m}07\pm69^{\rm s}$ \\
		  &					  &                                        &         &          &           & $-12\pm 2$  & LH  &14$^{\rm h}$02$^{\rm m}$08$\pm$90$^{\rm s}$
\enddata
\end{deluxetable*}

To better characterise M dwarf flares at meter wavelengths requires more observational time on individual sources to constrain flare rates. Also needed is more sensitive observations to investigate fine time-frequency structure of the flares. Simultaneous multi-wavelength observations would also add to this analysis. 

\acknowledgments
We thank R.~Wayth for helpful comments, and MWA Director S.~Tingay for awarding Director's Discretionary Time used for this project. We also thank the anonymous referee for their helpful discussion and comments. DLK was supported by NSF grant AST-1412421. TM acknowledges the support of the Australian Research Council through grant FT150100099.This scientific work makes use of the Murchison Radio-astronomy Observatory, operated by CSIRO. We acknowledge the Wajarri Yamatji people as the traditional owners of the Observatory site. Support for the operation of the MWA is provided by the Australian Government (NCRIS), under a contract to Curtin University administered by Astronomy Australia Limited. We acknowledge the Pawsey Supercomputing Centre which is supported by the Western Australian and Australian Governments. This research was conducted by the Australian Research Council Centre of Excellence for All-sky Astrophysics (CAASTRO), through project number CE110001020.  


\end{document}